\documentclass[useAMS,usenatbib]{mn2e}
\usepackage{graphicx}
\usepackage{amsmath}
\usepackage{amssymb}
\usepackage{times}
\newcommand{\hii}{{H{\scriptsize II} }}
\newcommand{\uchii}{{UCH{\scriptsize II} }}

\newcommand{\nhthree}{\mbox{NH$_3$}}
\newcommand{\kms}{\mbox{km\,s$^{-1}$}}

\title[New \nhthree~ masers towards NGC\,6334I]{New ammonia
masers towards NGC\,6334I} \author[A.J.Walsh et
al.]{A. J. Walsh$^{1}$, S. N. Longmore$^{2,3}$, S. Thorwirth$^4$, J. S. Urquhart$^5$,
C.R.Purcell$^6$\\ 
$^{1}$Centre for Astronomy, School of Maths, Physics and IT, James Cook University, Townsville, QLD 4811, Australia\\ 
$^{2}$School of Physics, University of New South Wales, Sydney, NSW 2052, Australia\\
$^{3}$Australia Telescope National Facility, CSIRO, Sydney, NSW 1710, Australia\\
$^{4}$Max-Planck-Institut f\"{u}r Radioastronomie, Auf dem H\"{u}gel 69, 53121, Bonn, Germany\\
$^{5}$School of Physics \& Astronomy, E.C. Stoner Building, The University of Leeds, Leeds LS2 9JT, UK\\
$^{6}$University of Manchester, Jodrell Bank Observatory, Macclesfield, Cheshire SK11 9DL, UK.\\
}
\begin{document}

\date{}


\maketitle

\begin{abstract}
We report the detection of new ammonia masers in the
non-metastable (8,6) and (11,9) transitions towards
the massive star forming region NGC\,6334\,I. Observations
were made with the ATCA interferometer and 
the emitting region appears unresolved in the 2.7\arcsec
$\times$ 0.8\arcsec~beam, with deconvolved sizes less
than an arcsecond. We estimate peak brightness
temperatures of $7.8 \times 10^5$ and $1.2 \times 10^5$\,K
for the (8,6) and (11,9) transitions, respectively. The masers appear
coincident both spatially and in velocity with a previously
detected ammonia (6,6) maser.
We also suggest that emission in the (10,9), (9,9)
and (7,6) transitions may also be masers, based on their narrow
line widths and overlapping velocity ranges with the above masers,
as observed with the single-dish Mopra radiotelescope.
 
\end{abstract}


\section{Introduction}
Ammonia (NH$_3$) is an extremely useful molecular tool for studying the interstellar
medium. Many inversion transitions occur in the easily observable 12\,mm
band. They include transitions from both metastable (J=K) and
non-metastable (J$>$K) levels of ammonia. Hyperfine structure in the
metastable transitions is commonly seen, which can be used to derive
optical depths, whereas comparison of different (J,K) transitions can
yield information on the rotational temperature (eg. \citealt{walmsley83,danby88}).

The first suggestion that ammonia may display maser action in
interstellar space was made by \citet{wilson82} who noted that (3,3)
emission toward W33 was not matched by (1,1), (2,2) and (4,4) transitions,
which were all in absorption. They suggested that this
may be due to a weak population inversion.
Maser emission in ammonia transitions was first unambiguously identified by
\citet{madden86} in the (9,6) and (6,3) transitions, shortly followed by
\citet{mauersberger86} in the (3,3) transition of $^{15}$NH$_3$.
In addition to these transitions, ammonia
masers have been detected in the following transitions:
(5,4) (7,5), (9,8), (10,8) \citep{mauersberger87},
(6,5) \citep{mauersberger88},
(5,5) \citep{cesaroni92},
(6,6) \citep{beuther07}
and (1,1) \citep{gaume96}.
$^{14}$NH$_3$ (3,3) masers were first detected by \citet{zhang95}.
Pumping mechanisms for most ammonia masers remain unclear. Metastable
transitions of ortho-ammonia, like (3,3) are thought to be collisionally
excited \citep{walmsley83}
but such a mechanism will only work for non-metastable transitions
where exceedingly high H$_2$ densities between $10^{10} - 10^{12}$ cm$^{-3}$
are found (eg. \citealt{madden86}). \citet{madden86} suggest
two alternative pumping mechanisms for non-metastable transitions:
either pumping by a strong infrared radiation field, such as
found around a deeply embedded high mass star, or by a fortuitous overlap
of a far-infrared line, which allows a population transfer from a (J,K)
inversion level to (J+1, K).

Ammonia masers are found in regions of high mass star formation, with
the best known example being W51 \citep{madden86,mauersberger87},
with other prominent examples being G9.62+0.19 \citep{cesaroni92,hofner94} and NGC6334I
\citep{kraemer95,beuther07}. NGC6334I forms the focus of this work. It is a nearby
(1.7kpc, \citealt{neckel78}) region of
high mass star formation, traced by bright infrared emission, an ultracompact (UC) \hii region,
mm continuum sources and methanol maser sites. See \citet{beuther07} for a 
more comprehensive summary of characteristics of the region.

\section{Observations and Data Reduction}
The initial observations were undertaken with the 22\,m Mopra
telescope near Siding Spring, Australia.
We obtained a spectrum of NGC\,6334I at 17 20 53.43, $-$35 47 2.2 (J2000)
using position switching on 2006 November 28$^{\rm th}$.
We used the new Mopra spectrometer (MOPS) in broadband
mode, which affords us four overlapping IFs of 2.2\,GHz each. Thus, we were
able to instantaneously cover 8\,GHz of frequency space between 19.5 and 27.5\,GHz.
Each IF has 8192 channels, resulting in a channel width of 269\,kHz, equivalent
to 4.1\kms~at 19.5\,GHz or 2.9\kms~at 27.5\,GHz. The observations consisted of
approximately 10 hours total integration (5 hours on source) with an rms
noise level of between 0.01 and 0.03\,K.

The follow-up observations were undertaken on 2007 May 3$^{\rm rd}$ on
the Australia Telescope Compact Array (ATCA) in Director's
time. The correlator setting was FULL\_256\_64\_256\_64, allowing
0.89\,\kms~velocity resolution. The primary beam was 2.3\arcmin~and the array
configuration was 1.5C, with baselines ranging from 77 to 4500\,m providing an
angular resolution of $\sim$1\arcsec. The two frequencies were tuned to cover the
ammonia (8,6) transition at 20.719221\,GHz and the (11,9) transition at 21.070739\,GHz.
These transitions were chosen for ATCA observations as they appeared the brightest
of previously undetected ammonia maser in the Mopra spectrum.

The region was
observed for 9$\times$20 minute cuts in each transition separated over
5 hours. A bright ($>$1 \,Jy), close ($<$5$^\circ$) phase calibrator
was observed for 3 minutes before and after each cut. ${\rm PKS~1921-293}$
and ${\rm PKS~1934-638}$ were used as the bandpass and primary
calibrator, respectively.

The data were reduced using the MIRIAD 
package. Bad visibilities were flagged, edge channels removed and the
gains/bandpass solutions from the calibrator were applied to the
visibilities. The data were Fourier transformed to form image cubes
with 0.2$\arcsec$ pixels using natural weighting. The images were
\emph{CLEAN}ed down to 3$\sigma$ above the noise to remove sidelobes
then convolved with a Gaussian beam of the synthesised beam size to
produce a final \emph{RESTOR}ed image. Continuum emission was
extracted from a low-order polynomial fit to line-free channels using
\emph{uvlin} and images were made from these visibilities in the same
way.
From previous observations of the primary calibrator, ${\rm
PKS~1934-638}$\footnote{see
http://www.narrabri.atnf.csiro.au/calibrators/}, errors in the
absolute flux scale are estimated to be $\sim$10\%. Noise levels were
6\,mJy/beam at the frequency of the (8,6) transition and 3.5\,mJy/beam
at the frequency of the (11,9) transition.


\section{Results and Discussion}
\begin{figure}
\includegraphics[width=0.5\textwidth]{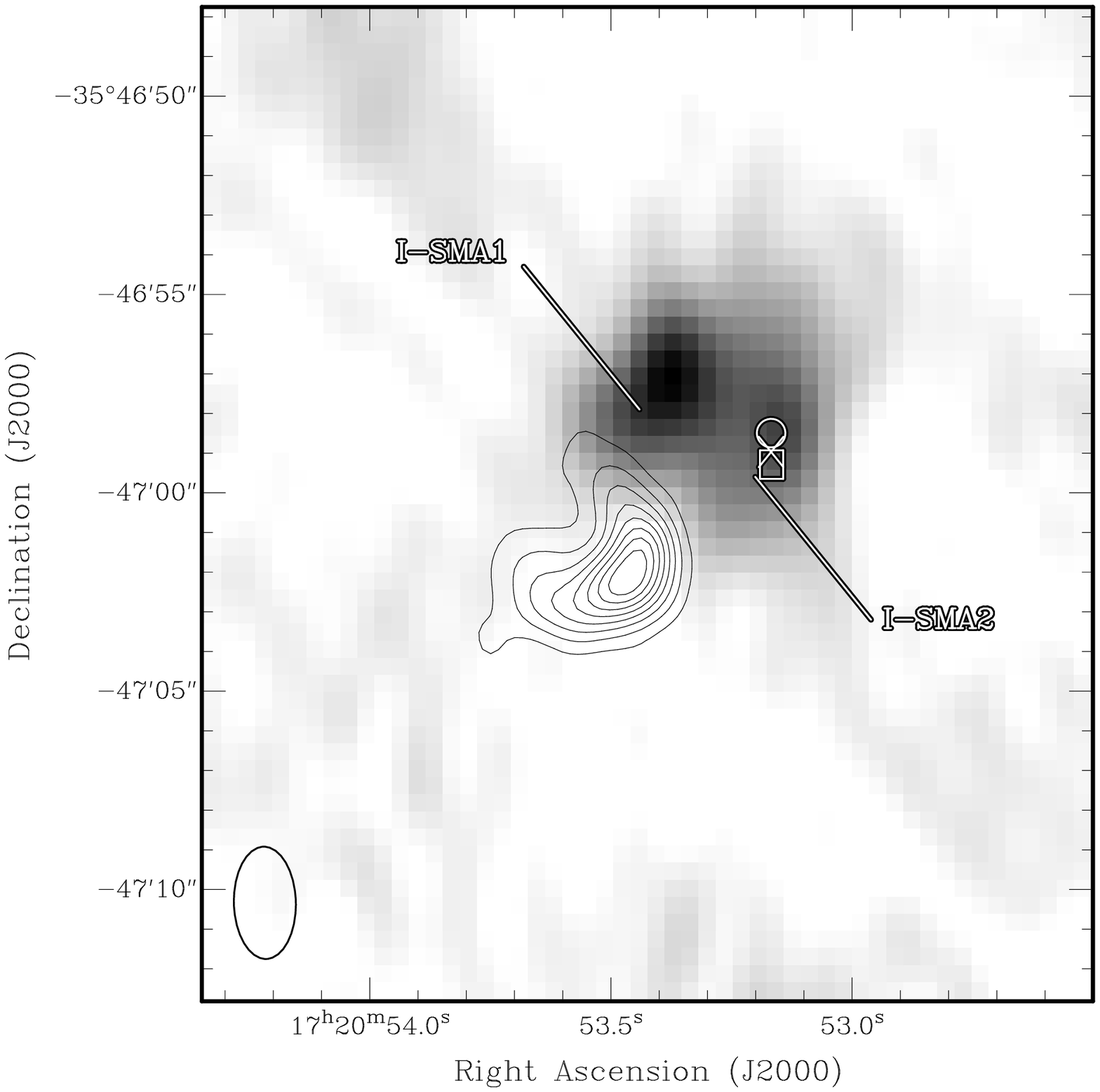}
\caption{Map of NGC\,6334I. The greyscale shows ammonia (4,4) integrated intensity thermal
emission. The two peaks of emission correspond to the mm continuum sources (I-SMA1 and I-SMA2)
detected by \citet{hunter06}. The contours represent 8.64\,GHz radio continuum emission from 
the \uchii region \citep{walsh98}. The ellipse in the bottom-left corner resresents
the FWHM beam for the ammonia (4,4) observations. The symbols represent the positions
of ammonia masers: the (8,6) maser is
the cross, the (11,9) maser is the circle and the (6,6) maser is the square \citep{beuther07}.}
\label{fig1}
\end{figure}

Figure \ref{fig1} shows radio continuum emission contours overlaid on thermal
ammonia (4,4) emission \citep{beuther07}. The (4,4) emission corresponds
well to two mm continuum sources I-SMA1 and I-SMA2 \citep{hunter06}. We find, in the ATCA observations,
that emission in both the (8,6) and (11,9) transitions is unresolved. We consider
these to be masers, for reasons given below. Because the emission is unresolved, 
we do not provide maps of each, but rather show symbols on Figure \ref{fig1}
to represent the positions of the maser emission.
Within positional errors (typically 1\arcsec~for the ATCA), all masers appear to be coincident.

\begin{figure}
\includegraphics[width=0.5\textwidth]{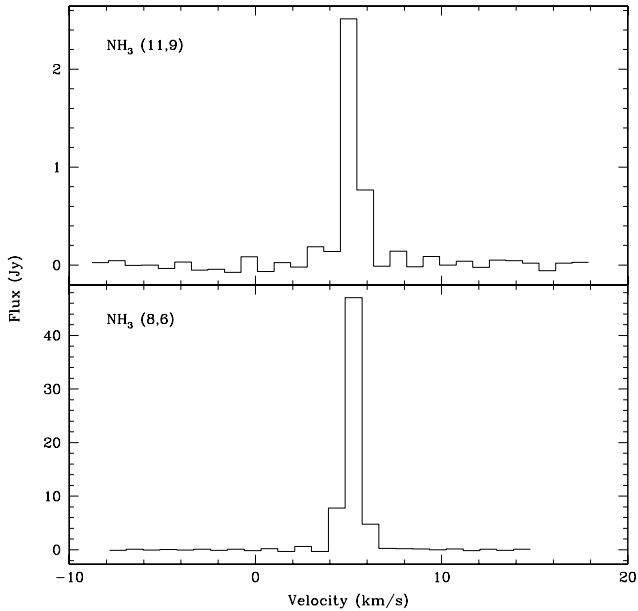}
\caption{Spectra for the (11,9) (top) and (8,6) (bottom) masers, as detected by the ATCA. The
velocity resolution is 0.89\,\kms.}
\label{fig2}
\end{figure}

Figure \ref{fig2} shows the spectra of each line. We measure fluxes of 53 and 3.5\,Jy
for the (8,6) and (11,9) transitions, respectively. These agree within errors to the respective
Mopra fluxes of 48\,Jy and 3.2\,Jy, implying there is no significant extended emission that might
have been missed by the ATCA obaservations. Therefore, we can be sure the emission comes from a
region no bigger than the ATCA beam of 2.7\arcsec $\times$ 0.8\arcsec.
We calculate peak brightness temperature lower limits based on equation \ref{eq1}
\begin{equation}
\frac{\rm K}{\rm Jy} = \frac{13.69 \times \lambda^2}{\theta_1 \times \theta_2}
\label{eq1}
\end{equation}
where $\lambda$ is the observing wavelength in millimetres and $\theta_1$ and $\theta_2$ are
the deconvolved major and minor axes in arcseconds, respectively. We find
the (8,6) emission has a deconvolved size of 0.9\arcsec $\times$ 0.2\arcsec and the (11,9)
emission has a deconvolved size of 0.7\arcsec $\times$ 0.1\arcsec. We therefore
calculate peak brightness temperatures
of 7.8 $\times 10^5$\,K for the (8,6) transition and 1.2 $\times 10^5$\,K for the (11,9) transition.
If the emission is thermal, it would require a local exciting source with a temperature in
excess of these values which we consider extremely unlikely.
Thus we interpret emission in both lines as masers.


\begin{figure}
\includegraphics[width=0.5\textwidth]{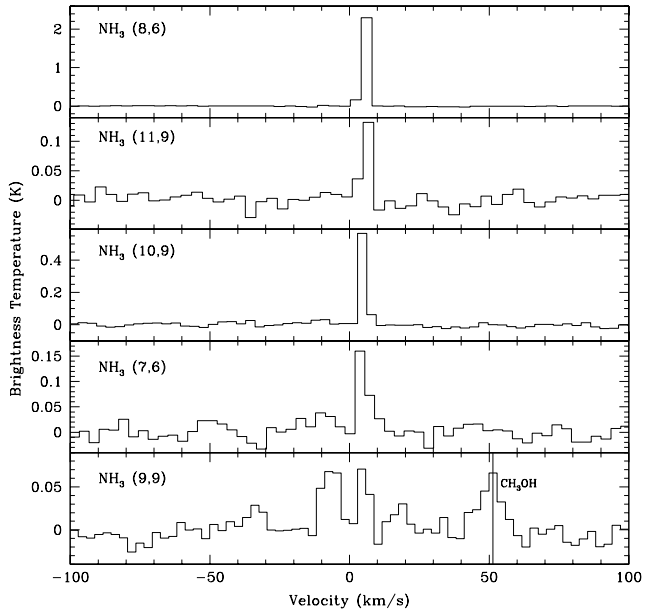}
\caption{Mopra spectra of ammonia transitions suspected to be masing. All spectra exhibit a narrow
emission feature between radial velocities 4.9 - 5.9\kms. Note that (9,9) emission in the bottom
panel exhibits the narrow maser-like feature within this velocity range as well as a broader
component at -6.7\kms, which is most likely due to thermal emission. The velocity of this thermal
component agrees well with velocities derived from other thermal lines. Note also that an unrelated spectral
feature appears in the (9,9) spectrum at 51\kms. This feature is due to E-type CH$_3$OH ($13_{2,11} - 13_{1,12}$).}
\label{fig3}
\end{figure}

Together with the (6,6) transition reported to be masing by \citet{beuther07},
we have identified three masing transitions of ortho-ammonia. Are there any other transitions
of ortho-ammonia that are masing? Our single-dish Mopra observations of NGC\,6334I suggest
there may be others. We detect narrow-lined emission in the (7,6), (10,9) and (9,9)
transitions, as shown in Figure \ref{fig3}.
Although the strengths of each of these lines is not enough to conclusively identify
them as masers, there is some circumstantial evidence that they may well be masers.
They all exhibit line widths less than 3.8\,\kms, which is close to the velocity
resolution of the Mopra observations. Indeed, the ATCA observations, with finer velocity resolution
show both the (8,6) and (11,9) lines to have line widths of 1.3\,\kms. In contrast to this,
all other detected ammonia lines in the Mopra spectrum have line widths between 5.8 and 7.3\,\kms,
and are clearly resolved in velocity.
Apart from their narrow line widths, emission in each of these transitions peaks around 5\,\kms,
as for the established (6,6), (8,6) and (11,9) masers. This is distinctively
different from thermal emission, which peaks between
-3 and -10\,\kms. Thermal emission is traced by ammonia (1,1), (2,2) and
CH$_3$OH (2$_{2,0}$ -- 2$_{1,1}$) \citep{beuther05}, as well as ammonia (3,3), (4,4), (5,5) and (6,6)
\citep{beuther07}. Note that the ammonia (6,6) observations identified both thermal and maser emission
that were clearly separated in velocity.

We note that the Mopra spectrum covers many other lines of ammonia between
19.5 and 27.5\,GHz and whilst some of these lines do show emission, it appears all others
exhibit broader emission features peaking around -3 and -10\,\kms~and so are almost certainly
thermal emission. We also note that all the potential masing transitions arise from ortho-ammonia.
It is not clear why we do not see any para-ammonia masers in NGC\,6334I.

Three methods for creating a population inversion of the ammonia upper levels are through collisional
excitation, excitation by infrared radiation or a fortuitous alignment of a molecular transition
that allows a (J,K) transition to populate the (J+1,K) upper level \citep{madden86}. It is unlikely
that collisional excitation is at work here because extremely high densities ($10^{10} - 10^{12} {\rm cm}^{-3}$)
are required. Since we see at least three masing transitions, and perhaps six, it is unlikely
this is due to many lucky alignments of molecular transitions allowing population
of the non-metastable states. Therefore, we believe the pumping mechanism for ammonia masers
in NGC\,6334I must be via infrared photons. This case is strengthened as NGC\,6334I is known to
be one of the brightest infrared sources in our sky, being intrinsically luminous and relatively
nearby (1.7\,kpc). This may also explain why these new masers only show thermal emission in other
regions.

The detection of new ammonia masers in NGC\,6334I indicates that this source
is fertile ground for ammonia maser research.
Confirmation (or otherwise) of the potential (7,6), (10,9) and (9,9) masers, as well as searching
for other non-metastable transitions of ortho-ammonia may yield more maser detections, which will
greatly increase the chances of developing theoretical models for ammonia maser pumping and hopefully
will lead to a better understanding of NGC\,6334I, which gives rise to these rare masers.
 
\section{Conclusions}

We have observed the two non-metastable transitions of ammonia: (8,6) and (11,9), using the ATCA.
The (8,6) transition has a peak brightness temperature of 7.8 $\times 10^5$\,K, whereas the (11,9)
transition has a peak brightness temperature of 1.2 $\times 10^5$\,K, indicating that both are
masers. The (11,9) transition is E$_{\rm l}/k=1449$\,K above ground making it the
highest energy ammonia maser currently known.
The position of both masers is consistent with being coincident, to within positional uncertainties,
with maser emission in the previously detected (6,6) transition \citep{beuther07}.

Single-dish observations of (7,6), (9,9) and (10,9) suggest they may also be masers based on their
narrow line widths and overlapping velocity range with the above-mentioned masers, which is distinctly
different from the thermal emission systemic velocity. High spatial resolution observations of these
transitions are planned to decide whether or not these are masers as well.

\section{Acknowledgments}
The authors would like to thank Christian Henkel for useful discussions
related to this work. We would also like to thank Henrik Beuther for provision of ammonia
and continuum data used in this work. We would also like to thank Paul Ho, the referee,
who has greatly helped improve the quality of this work.
SNL is supported by a scholarship from the School of Physics at
UNSW. The Australia Telescope is funded by the Commonwealth of
Australia for operation as a National Facility managed by CSIRO. This
research has made use of NASA's Astrophysics Data System.


\begin{thebibliography}{}
\bibitem[\protect\citeauthoryear{Beuther et al.}{2005}]{beuther05} Beuther, H., Thorwirth, S., Zhang, Q.,
Hunter, T. R., Megeath, S. T., Walsh, A. J. \& Menten, K. M. 2005, ApJ, 627, 834
\bibitem[\protect\citeauthoryear{Beuther et al.}{2007}]{beuther07} Beuther, H., Walsh, A. J.,
Thorwirth, S., Zhang, Q., Hunter, T. R., Megeath, S. T., Menten, K. M. 2007, A\&A, 466, 989
\bibitem[\protect\citeauthoryear{Cesaroni et al.}{1992}]{cesaroni92} Cesaroni, R., Walmsley, C. M.,
Churchwell, E. 1992, A\&A, 256, 618
\bibitem[\protect\citeauthoryear{Danby et al.}{1988}]{danby88} Danby, G., Flower, D. R., 
Valiron, P., Schilke, P., Walmsley, C. M. 1988, MNRAS, 235, 229
\bibitem[\protect\citeauthoryear{Gaume et al.}{1996}]{gaume96} Gaume, R. A.,
Wilson, T. L., Johnston, K. J. 1996, ApJL, 457, 47
\bibitem[\protect\citeauthoryear{Hofner et al.}{1994}]{hofner94} Hofner, P.,
Kurtz, S., Churchwell, E., Walmsley, C. M., Cesaroni, R. 1994, 1994, ApJL, 429, 85
\bibitem[\protect\citeauthoryear{Hunter et al.}{2006}]{hunter06} Hunter, T. R.,
Brogan, C. L., Megeath, S. T., Menten, K. M., Beuther, H. \& Thorwirth, S. 2006, ApJ, 649, 888
\bibitem[\protect\citeauthoryear{Kraemer \& Jackson}{1995}]{kraemer95} Kraemer, K. E., Jackson, J. M.
1995, ApJL, 439, 9
\bibitem[\protect\citeauthoryear{Madden et al.}{1986}]{madden86} Madden, S. C., Irvine, W. M.,
Matthews, H. E., Brown, R. D., Godfrey, P. D. 1986, ApJL, 300, 79
\bibitem[\protect\citeauthoryear{Mauersberger et al.}{1986}]{mauersberger86} Mauersberger, R.,
Wilson, T. L., Henkel, C. 1986, A\&AL, 160, 13
\bibitem[\protect\citeauthoryear{Mauersberger et al.}{1987}]{mauersberger87} Mauersberger, R.,
Henkel, C., Wilson, T. L. 1987, A\&A, 173, 352
\bibitem[\protect\citeauthoryear{Mauersberger et al.}{1988}]{mauersberger88} Mauersberger, R.,
Wilson, T. L., Henkel, C. 1988, A\&A, 201, 123
\bibitem[\protect\citeauthoryear{Neckel}{1978}]{neckel78} Neckel, T. 1978, A\&A, 69, 51
\bibitem[\protect\citeauthoryear{Walmsley \& Ungerechts}{1983}]{walmsley83} Walmsley, C. M.,
Ungerechts, H. 1983, A\&A, 122, 10
\bibitem[\protect\citeauthoryear{Walsh et al.}{1998}]{walsh98} Walsh, A. J.,
Burton, M. G., Hyland, A. R., Robinson, G., 1998, MNRAS, 301, 640
\bibitem[\protect\citeauthoryear{Wilson et al.}{1982}]{wilson82} Wilson, T. L., Batrla, W., Pauls, T. A.
1982, A\&AL, 110, 20
\bibitem[\protect\citeauthoryear{Zhang \& Ho}{1995}]{zhang95} Zhang, Q. \& Ho, P. T. P. 1995, ApJL, 450, 63
\end{thebibliography}
\end{document}